\documentclass[prb, twocolumn, showpacs, superscriptaddress, longbibliography, amsmath, amssymb]{revtex4-1}

\usepackage{graphicx}
\usepackage{dcolumn}
\usepackage{bm}
\usepackage[dvips]{color}

\newcommand{\vect}[1]{\mathbf{#1}}

\begin{document}
\preprint{APS/123-QED}

\title{Multiferroicity of CuCrO$_2$ tested by ESR}

\author{S.K. Gotovko}
\affiliation{P.L. Kapitza Institute for Physical Problems, RAS,  Moscow 119334, Russia}
\affiliation{Moscow Institute of Physics and Technology,  Moscow 117303, Russia}

\author{T.A. Soldatov}
\affiliation{P.L. Kapitza Institute for Physical Problems, RAS,  Moscow 119334, Russia}
\affiliation{Moscow Institute of Physics and Technology,  Moscow 117303, Russia}

\author{L.E. Svistov}
\email{svistov@kapitza.ras.ru}
\affiliation{P.L. Kapitza Institute for
Physical Problems, RAS,  Moscow 119334, Russia}

\author{H.D. Zhou}
\affiliation{Department of Physics and Astronomy, University of
Tennessee, Knoxville, Tennessee 37996, USA}

\date{\today}

\begin{abstract}
We have carried out the ESR study of the multiferroic triangular antiferromagnet CuCrO$_2$ in the presence of an electric field. The shift of ESR spectra by the electric field was observed; the observed value of the shift exceeds that one in materials with linear magnetoelectric coupling. It was shown that the low-frequency dynamics of magnetically ordered CuCrO$_2$ is defined by joint oscillations of the spin plane and electric polarization. The results demonstrate qualitative and quantitative agreement with theoretical expectations of a phenomenological model (V.~I.~Marchenko (2014)).
\end{abstract}


\maketitle

\section{Introduction}

CuCrO$_2$ is an example of a quasi-2D antiferromagnet ($S=3/2$ and $T_c\approx 25$~K~\cite{Kadowaki_1990}) with a triangular lattice structure. The neutron scattering investigations in CuCrO$_2$ single crystals detected a 3D planar magnetic order with an incommensurate wave vector that slightly differs from the wave vector of a commensurate 120-degree structure.~\cite{Poienar_2009} The magnetic ordering is accompanied by a simultaneous crystallographic distortion of the regular triangular lattice and by the appearance of electric polarization.~\cite{Kimura_JPSJ_2009}
According to the recent experimental study of the electric polarization and NMR on nonmagnetic ions, CuCrO$_2$ demonstrates a rich magnetic phase diagram.~\cite{Zapf_2014, Sakhratov_2016, Portugall_2017} Some of these phases were identified as a nematic (often termed ``chiral" phase) characterized by the absence of dipolar order and by the presence of spontaneous electric polarization.
This observation has stimulated our interest in the nature of the coupling between magnetic order and spontaneous electric polarization (termed multiferroicity) in this magnet.

We present an ESR study of CuCrO$_2$ in fields much below saturation ($\mu_0 H_{sat}\approx 280 $~T) in the presence of an electric field. Uniform excitations in the magnetically ordered state tested by ESR can be regarded as oscillations of the spin plane of magnetic structure. The orientation of the spin plane defines the electric polarization direction.~\cite{Kimura_PRL_2009} Therefore, the oscillation of spin plane should be accompanied by oscillation of the electric polarization direction. Such oscillations could be excited by an alternating electric field and the ESR frequencies should be dependent on the applied electric field. It is expected that the influence of the electric field on the resonance frequencies in the case of a multiferroic material  can be stronger than in the case of material with the lineal magnetoelectric coupling (Refs.~\onlinecite{Smirnov_1994},~\onlinecite{Shengelaya_2012}), because the value of spontaneous electric polarization in multiferroic exceeds the value induced by magnetic fields of resonable magnitudes. Results of the ESR study of CuCrO$_2$ in an electric field are reported here. The obtained results are compared with the phenomenological theory of magnetic properties of CuCrO$_2$ developed in Ref.~\onlinecite{Marchenko_2014}.

\section{Crystal and magnetic structure}

\begin{figure}[b!]
\includegraphics[width=0.95\columnwidth,angle=0,clip]{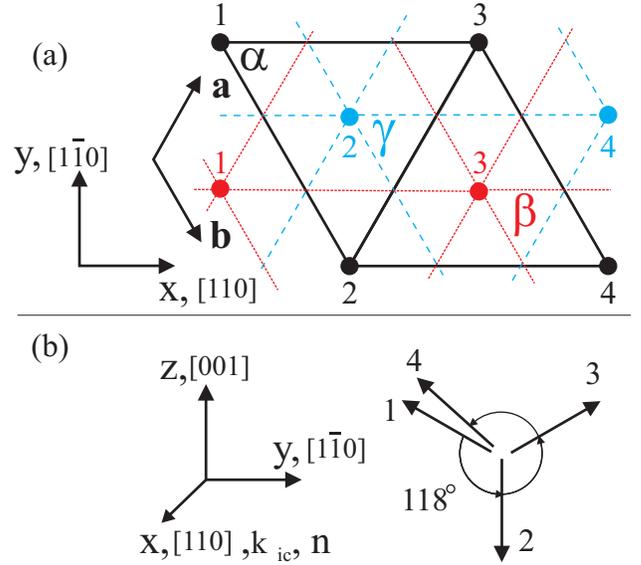}
\caption{(color online) (a) Crystal structure of CuCrO$_2$ projected on the
$ab$-plane. The three layers, $\alpha, \beta, \gamma$, are the positions of
Cr$^{3+}$ ions. (b) Scheme of the spin structure at zero magnetic field for $\vect{k}_{ic}\parallel [110]$ (the spins with the same numbers in (a) are codirected).
}
\label{fig:structure}
\end{figure}

The CuCrO$_2$ structure  consists of magnetic Cr$^{3+}$ (3$d^3$, $S=3/2$), nonmagnetic Cu$^+$, and O$^{2-}$ triangular lattice planes, which are stacked along the $c$-axis in the sequence Cr--O--Cu--O--Cr (space group $R\bar{3}m$, $a=2.98$~\AA{} and $c=17.11$~\AA{} at room temperature).~\cite{Poienar_2009}

The positions of magnetic Cr$^{3+}$ in the crystal structure of CuCrO$_2$ projected on the $ab$-plane are shown in Fig.~\ref{fig:structure}(a). The ions of different triangular planes $\alpha\gamma\beta$ separated from each other by $c/3$ are marked with different colors.

In the magnetically ordered state, the triangular lattice is distorted such that one side of the triangle becomes slightly smaller than the other two sides: $\Delta a / a \simeq 10^{-4}$.~\cite{Kimura_JPSJ_2009}

The magnetic structure of CuCrO$_2$ has been intensively investigated by neutron diffraction experiments.~\cite{Poienar_2009, Kadowaki_1990, Soda_2009, Soda_2010, Frontzek_2012} It was found that the magnetic ordering in CuCrO$_2$ occurs in two stages.~\cite{Frontzek_2012, Aktas_2013} At the higher transition temperature $T_{c1}=24.2$~K, a transition to a 2D ordered state occurs, whereas below $T_{c2}=23.6$~K, a 3D magnetic order with the incommensurate propagation vector $\vect{k}_{ic}= (0.329,0.329,0)$ along the distorted side of triangular lattice planes is established.~\cite{Kimura_JPSJ_2009} The magnetic moments of Cr$^{3+}$ ions can be described by the expression
\begin{eqnarray}
\vect{M}(\vect{r}_{i,j})=M_1\vect{e}_1\cos(\vect{k}_{ic}\vect{r}_{i,j}+\theta)+M_2\vect{e}_2\sin(\vect{k}_{ic}\vect{r}_{i,j}+\theta),
\label{eqn:spiral}
\end{eqnarray}
where $\vect{e}_1$ and $\vect{e}_2$ are two perpendicular unit vectors determining the spin plane orientation with the normal vector $\vect{n}=\vect{e}_1 \times \vect{e}_2$, $\vect{r}_{i,j}$ is the vector to the ($i,j$)-th magnetic ion, and $\theta$ is an arbitrary phase. The spin plane orientation and the propagation vector of the magnetic structure are schematically shown at the bottom of Fig.~\ref{fig:structure}. For zero magnetic field, $\vect{e}_1$ is parallel to $[1\bar{1}0]$ with $M_1 = 2.2(2)~\mu_B$, while $\vect{e}_2$ is parallel to $[001]$ with $M_2 = 2.8(2)~\mu_B$.~\cite{Frontzek_2012} The pitch angle between the neighboring Cr$^{3+}$ moments corresponding to the observed value of $\vect{k}_{ic}$ along the distorted side of triangular lattice planes is equal to 118.5$^\circ$, which is very close to the 120$^\circ$ expected for a regular triangular lattice plane structure.

Owing to the crystallographic symmetry at $T > T_c$ in the ordered phase ($T < T_c$), we can expect {\it six} magnetic domains. The propagation vector of each domain can be directed along one side of the triangle and can be positive or negative. As reported in Refs.~\onlinecite{Soda_2010, Svistov_2013, Sakhratov_2014}, the distribution of the domains is strongly affected by the cooling history of the sample.

Inelastic neutron scattering experiments have shown that CuCrO$_2$ can be considered as a quasi-2D magnet.~\cite{Poienar_2010} The spiral magnetic structure is defined by the strong exchange interaction between the nearest Cr$^{3+}$ ions within the triangular lattice planes with the exchange constant $J_{ab}=2.3$~meV. The inter-planar interactions are approximately 20 times weaker than the in-plane interaction and are frustrated.

Simultaneously with the appearance of three-dimensional magnetic order, the sample acquires an electric polarization, which is governed by the magnetic structure of CuCrO$_2$.

Results of the magnetization, electric polarization and ESR and NMR experiments~\cite{Kimura_PRL_2009, Svistov_2013, Sakhratov_2014} have been discussed within the framework of the planar spiral spin structure at fields studied experimentally: $\mu_0H < 14$~T~$\ll \mu_0H_{sat}$ ($\mu_0H_{sat}\approx 280$~T). The orientation of the spin plane is determined by the biaxial crystal anisotropy and the applied magnetic and electric fields.

According to Ref.~\onlinecite{Marchenko_2014}, the main properties of antiferromagnetic CuCrO$_2$ have a natural explanation based on the Dzyaloshinski--Landau theory of magnetic phase transitions.

The consideration of exchange interactions demonstrates that the crystal structure of CuCrO$_2$ allows the Lifshitz invariant that couples the spins of neighboring triangular planes; this explains the helicoidal spin structure with an incommensurate wave vector. The proximity of the wave vector of the magnetic structure for CuCrO$_2$ (0.329,0.329,0) to the wave vector of a simple 120-degree structure (1/3,1/3,0) demonstrates the smallness of the Lifshitz invariant compared with the intraplane exchange interaction. Symmetry analysis of relativistic interactions in CuCrO$_2$~\cite{Marchenko_2014} explains the experimentally observed  magnetic anisotropy and the electric polarization codirected with the vector $\boldsymbol{n}$ of the magnetic structure.

Using the notation in Ref.~\onlinecite{Marchenko_2014}, the energy of CuCrO$_2$ dependent on the spin plane orientation with respect to crystallographic axes and the applied magnetic and electric fields can be written as

\begin{eqnarray}
U=\frac{\beta_1}{2}n^2_z+\frac{\beta_2}{2}n^2_y-\frac{\chi_{||}-\chi_\perp}{2}(\textbf{n}\textbf{H})^2 \nonumber\\-\lambda_\perp (n_xE_x +n_yE_y)-\lambda_\parallel n_zE_z
\label{eqn:energy}
\end{eqnarray}
The first two terms describe the anisotropy energy. One {\it hard} axis for the normal vector
$\vect{n}$ is parallel to the $\vect{z}$ direction and the second axis $\vect{y}$ is
perpendicular to the direction of the distorted side of the triangle ($\beta_1, \beta_2>0$). The directions of $\vect{x}$,$\vect{y}$,$\vect{z}$ axes are shown in Fig.~\ref{fig:structure}.  The anisotropy along the $c$ direction dominates with the anisotropy constant approximately
a hundred times larger than that within the $ab$ plane~\cite{Svistov_2013}: $\beta_1=355$~kJ/m$^3$ and $\beta_2=3.05$~kJ/m$^3$.
A magnetic phase transition was observed for the
field applied perpendicular to one side of the triangle
($\vect{H}\parallel [1\bar{1}0]$) at $\mu_0H_c \approx 5.5$~T, which was consistently
described~\cite{Kimura_PRL_2009, Soda_2010, Svistov_2013} by the transition
of the spin plane from $(110)$ ($\vect{n}\perp\vect{H}$) to $(1\bar{1}0)$
$(\vect{n}\parallel\vect{H})$. This spin transition to an ``umbrella like'' phase
occurs due to the weak susceptibility anisotropy of the spin structure ($\chi_{\parallel}\approx 1.045\chi_{\perp}$), where $\parallel$ and $\perp$
refer to fields parallel and perpendicular to
$\vect{n}$. The transition field is defined by  $H_c^2 = \beta_2/(\chi_\parallel - \chi_\perp)$. The third term in Eq.~(\ref{eqn:energy}) takes the anisotropy of magnetic susceptibility into account. The experimental value of the susceptibility is $\chi_{\perp}\approx 2400$~J/T$^2$m$^3$.~\cite{Kimura_PRB_2008, Zapf_2014} The last two terms describe the interaction of the spontaneous electric polarization caused by magnetic ordering $\vect{p}=(\lambda_{\perp}n_x,\lambda{_\perp}n_y,\lambda_{\parallel} n_z$) with the external electric field $(E_x,E_y,E_z)$. The experimental study of electric polarization gives the value of $\lambda{_\perp}$ as 120$\div$130~$\mu$C/m$^2$.~\cite{Kimura_PRB_2008}
Comparing values of interactions of the spin structure with the crystal environment (the first two terms of the equation) and with the magnetic and electric fields, we can conclude that the interaction with the electric field in all the experimentally studied range of $E$ is small. For example, the spin flop transition in the electric field predicted in Ref.~\onlinecite{Marchenko_2014} is expected at the applied field $E_c\approx 30000$~kV/m. Fields used in our experiments were much smaller: $E < 1000$~ kV/m.

The magnetic structure of CuCrO$_2$ without the electric field was studied previously with the ESR technique for  magnetic fields applied along rational crystal axes: $\vect{H}\parallel [110],~[1\bar{1} 0]$.~\cite{Yamaguchi_2010, Svistov_2013}
Three antiferromagnetic resonance frequencies are expected for the planar spiral spin structure. One resonance frequency for an incommensurate planar structure is expected to be zero: $\nu_1(H)=0$. This is a consequence of the invariability of the spin structure energy under its rotation around the vector $\vect{n}$ of the spin plane. Two other eigenfrequencies correspond to oscillations around $\vect{y}$ and $\vect{z}$: $\nu_2(H=0)=\gamma\sqrt{\beta_1/\chi_\perp}\approx 340$~GHz and $\nu_3(H=0)=\gamma\sqrt{\beta_2/\chi_\perp}\approx 31.5$~GHz.~\cite{Svistov_2013} Here, $\gamma=hg\mu_B=28$~GHz/T, the g-factor is equal to 2, and h is Planck's constant.

\begin{figure}[h]
\includegraphics[width=0.95\columnwidth,angle=0,clip]{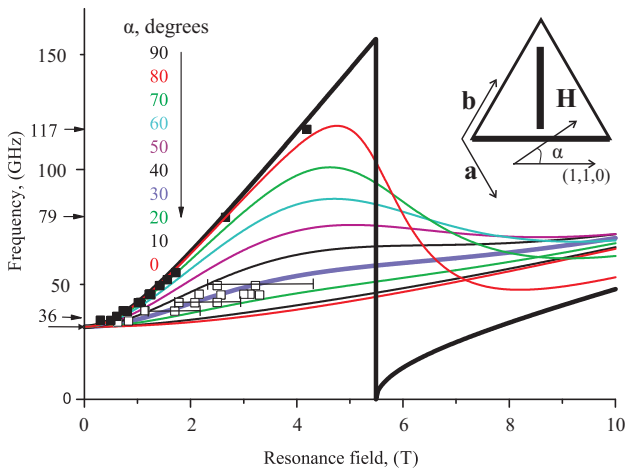}
\caption{(color online) Lines show frequency--resonance-field diagrams ($\nu_3(H_R)$) computed for different angles $\alpha$ between $\vect{H}$ and [110]. Symbols show $\nu (H_R)$ measured at the field perpendicular to one side of the triangular structure. Solid symbols correspond to absorption lines from domain ``A" ($\alpha=90^{\circ}$) and  open symbols - from domains ``B" and ``C" ($\alpha=\pm 30^{\circ}$). $T=4.2$~K.
}
\label{fig:angle2}
\end{figure}

Frequency--resonance-field diagrams ($\nu_3(H_R)$) computed for the model described by Eq.~(\ref{eqn:energy}) are shown in Fig.~\ref{fig:angle2}. These dependencies for the low-frequency branch $\nu_3$ were  computed for different field directions in the $ab$ plane  using the parameters $\chi_{\parallel}/\chi_{\perp}=1.042, H_c=5.5$~T and $\nu_3(0)=31.6$~GHz. The values of these parameters are in agreement with parameters  defining the anisotropic part of energy (Eq.~(\ref{eqn:energy})). For the computations, we used the theory of spin dynamics for magnets with dominant exchange interactions~\cite{Andreev_1980}. The application of this theory to the coplanar magnetic structures was described in Refs.~\onlinecite{Prozorova_1985,  Zaliznyak_1988, Glazkov_2016}.

The $\nu_3(H_R)$ dependence for $\vect{H} \parallel [1\bar{1}0]$
demonstrates an abrupt jump at the spin-flop field $H_c$. For fields much larger than $H_c$ the dependences asymptotically tend to linear, with the slope defined by anisotropy of the spin structure susceptibility $\gamma H \sqrt{\chi_{\parallel}/\chi_{\perp}-1}$.

Experimental values of resonance fields $H_R$ of the absorption lines measured at different frequencies are plotted in the same figure (Fig.~\ref{fig:angle2}) with symbols. The magnetic field $\vect{H}$ was applied perpendicularly to one side of the triangle. This field direction was used in ESR experiments in the electric field (see Fig.~\ref{fig:setup}). Symbols showing $H_R$ by solid squares correspond to absorption lines from the domain ``A", where $\vect{H} \parallel [1 \bar{1} 0]$. Open squares correspond to absorption lines from two other domains, ``B" ($\vect{H}\parallel[2 1 0]$) and ``C" ($\vect{H}\parallel[\bar{1}\bar{2}0]$). The points are in agreement with the theoretical expectations (black and blue bold lines). At this field direction, absorption lines from domain ``A" and domains ``B" and ``C" are well separated, which simplifies the interpretation of the results.

\section{Sample preparation and experimental details}

\begin{figure}[h]
\includegraphics[width=0.95\columnwidth,angle=0,clip]{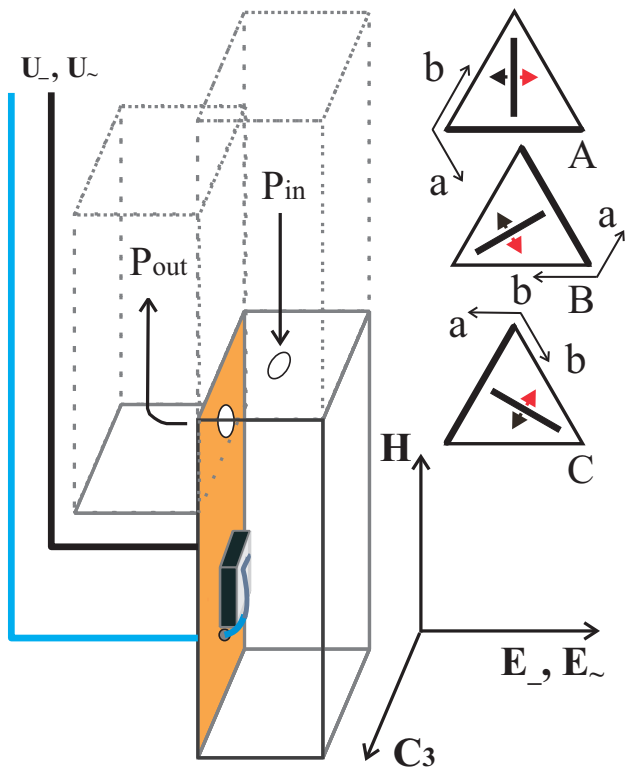}
\caption{(color online) Left panel: Scheme of the experimental cell: a rectangular resonator with holes of coupling with the UHF transmission line, the sample of a plate form with a silver paste electrode. Right panel: Three possible magnetic domains in the sample, shown schematically as ``A", ``B", ``C", and the mutual orientation of the applied electric and magnetic fields and crystallographic axes of the sample. The bold lines on the scheme aligned along heights of the triangles indicate the projection of spin planes at $H=0$. The bold side of the triangle marks the direction of the wave vector of the magnetic structure in each domain. The developed plane of the sample is perpendicular to one side of the triangular crystal structure. For domain ``A", $\vect{H}\parallel[1 \bar{1} 0]$ and $\vect{E}\parallel[110]$;  for ``B", $\vect{H}\parallel[2 1 0]$ and $\vect{E}\parallel[0\bar{1}0]$; and for ``C", $\vect{H}\parallel[\bar{1}\bar{2}0]$ and $\vect{E}\parallel[\bar{1}00]$.
Red and black arrows indicate the two possible directions of electric polarization for domains ``A", ``B" and ``C".
}
\label{fig:setup}
\end{figure}

The CuCrO$_2$ samples used were from the same growth butch as described in Refs.~\onlinecite{Sakhratov_2014,  Sakhratov_2016}. Samples were sawn in  plates 0.31~mm in thickness, with an $\approx 1\times 3$~mm$^2$ of plane size. The plane of the samples was perpendicular to one side of the triangle of the crystal structure. The sample was glued on the wall of a high-frequency resonator of transmission type. The wall of the rectangular resonator was used as one electrode. The opposite plane of the sample plate was covered by silver paste and used as another electrode. The scheme of the low-temperature part of the experimental cell, which allows conducting the ESR experiments in an electric field, is shown in the left panel of Fig.~\ref{fig:setup}. The mutual orientation of the applied electric and magnetic fields and crystallographic axes of the sample is shown in the right panel of the figure.

The shift of absorption lines by a permanent electric field $E\_$ was not sufficient for its reliable observation, and therefore the influence of the electric field on absorption was studied by the modulation method. In the described experiments, the electric field $E_{\sim}$ was oscillating instead of the oscillating magnetic field used in traditional EPR experiments. Such a technique was used previously in Refs.~\onlinecite{Smirnov_1994}, \onlinecite{Shengelaya_2012}. The oscillation frequency of the alternating electric field was in the range 100--300 Hz. The experimental results did not depend on the modulation frequency.

In addition to $E_{\sim}$, the permanent electric field $E\_$ was applied to the sample for its electric polarization. After application of the electric field of a sufficient magnitude, only energetically favorable electric domains are expected in the sample.~\cite{Kimura_PRL_2009} The electric polarizations of these domains are shown in Fig.~\ref{fig:setup} with red arrows. The amplitude of the alternating electric field $E_{\sim}$ was smaller than the permanent electric field $E\_$ to avoid electric depolarization of the sample. The magnetic field dependences of both the high-frequency power $P_{tr}(H)$ transmitted through the resonator and the amplitude of its oscillation at the frequency of the applied alternating electric field $P_{tr}^{\sim}(H)$ were measured. The knowledge of these two parameters allows determining the dependence of the frequency of uniform magnetic oscillations on the magnitude of the electric field and comparing it with the theory. The theoretical expectation essentially depends on the form of the anisotropic energy in Eq.~(\ref{eqn:energy}). To verify the validity of this representation, the angular dependences of resonance fields in the magneto-ordered state of CuCrO$_2$ were measured. The results of this study, which are in agreement with the model (Eq.~(\ref{eqn:energy})), are given in an appendix.

\section{Experimental results}

\begin{figure}[h]
\includegraphics[width=0.95\columnwidth,angle=0,clip]{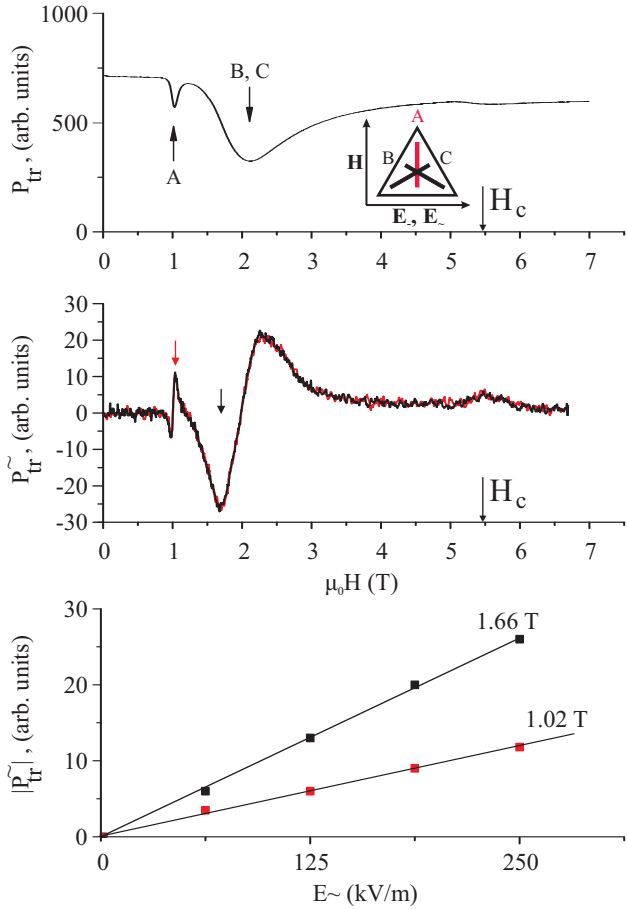}
\caption{(color online) Upper panel: Field scan of the transmitted power. Middle panel: Field scans of oscillation amplitude of the transmitted HF power $P^{\sim}_{tr}$ at the frequency of the applied alternating electric field $E^{\sim}$ measured with the phase detection technique.  $E_{\_}=+500$~kV/m and $E^{\sim}=250$~kV/m. Red and black scans correspond to different polarites of the superconducting solenoid. The arrows at the abscissa axes show the field of spin-flop reorientation $H_c$ in domain ``A". Bottom panel: $|P^{\sim}_{tr}(E^{\sim})|$ dependences measured at $H=1.02, 1.66$~T shown in the middle panel with arrows. $\nu=42.2$~GHz and $T=4.2$~K.
}
\label{fig:line1}
\end{figure}

The dependence of the high-frequency power $P_{tr}$ transmitted through the resonator on the applied static field $H$ measured at $T=4.2$~K and $\nu=42.2$~GHz is shown in the upper panel of Fig.~\ref{fig:line1}. The low-field absorption line corresponds to ESR from domain ``A", and the broad absorption line at high fields corresponds to ESR from domains ``B" and ``C". The peculiarity at $\mu_0 H_c=5.5$~T corresponds to spin-flop reorientation in domain ``A". The permanent electric field  was applied to the sample to polarize it.  The amplitude of the oscillations of the transmitted high-frequency power $P^{\sim}_{tr}$ at the frequency of the applied alternating electric field measured with the phase detection technique is shown in the middle panel. The reference signal was in phase with the alternating electric field $E^{\sim}$ applied to the sample. The positive sign of $P^{\sim}_{tr}$ corresponds to oscillations of $P_{tr}$ in phase with $E^{\sim}$, whereas the negative sign corresponds to out-of-phase oscillations. The values of both $P_{tr}$ and $P^{\sim}_{tr}$ are presented in arbitrary, but the same units. The red and black curves show $P^{\sim}_{tr}(H)$ measured at two opposite directions of the field $\vect{H}$. $P^{\sim}_{tr}$ is independent on the polarity of the static magnetic field within the precision of the experiment. The bottom panel of Fig.~\ref{fig:line1} demonstrates the linearity of  $P^{\sim}_{tr}$ on $E^{\sim}$ observed in the range $|E^{\sim}|<|E_{\_}|$.

\begin{figure}[h]
\includegraphics[width=0.95\columnwidth,angle=0,clip]{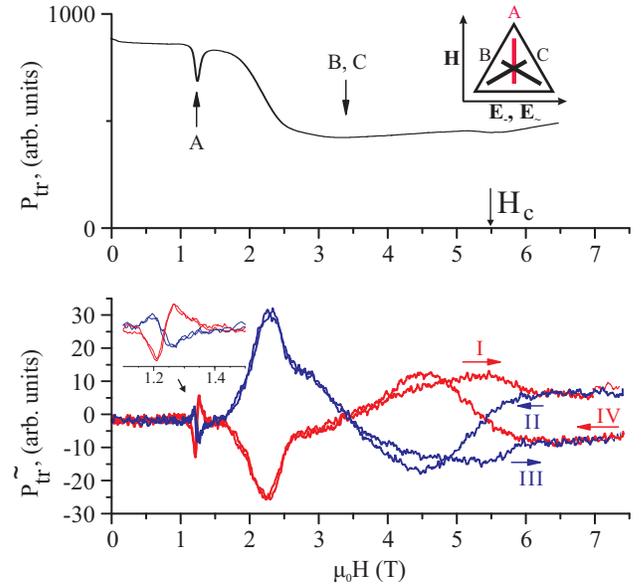}
\caption{(color online) Upper panel: Field scan of $P_{tr}$. Bottom panel: four consequent field scans of $P^{\sim}_{tr}$ measured at $E^{\sim}=250$~kV/m. Scans I and IV were measured  at $E_{\_}=+500$~kV/m (red lines) and II and III were measured at $E_{\_}=-500$~kV/m (blue lines). Directions of field scans are shown with arrows. The scoped scans of  $P^{\sim}_{tr}(H)$ in the range of the absorption line from domain ``A" are shown in the inset. $\nu=45.6$~GHz and $T=4.2$~K.
}
\label{fig:line2}
\end{figure}

The scans of $P_{tr}(H)$ and  $P^{\sim}_{tr}(H)$ measured at  $E_{\_}$ of different signs are shown in Fig.~\ref{fig:line2} in the upper and bottom panel. The used value of the permanent field $E_{\_}=\pm 500$~kV/m was sufficient for electric polarization of the sample at low magnetic fields. Llines I and IV in the bottom panel were measured at the positive sign of $E_{\_}$ at the increase and decrease in the static field $H$ respectively. Lines II and III were measured at the negative sign of the permanent electric field. The scan directions are shown with arrows in the figure. The sign of the electric field $E_{\_}$ was switched at  $H=7.5$~T. This figure demonstrates that the sign of $P^{\sim}_{tr}(H)$ is defined by the electric polarization of the sample. The observed $P^{\sim}_{tr}(H)$ corresponds to the shift  of the absorption line to lower fields in the time intervals when the oscillating field  is codirected with the polarization of the sample, and to higher fields in the time intervals when $\vect{E}^{\sim}$ is antiparallel to the polarization of the sample. The hysteresis of $P^{\sim}_{tr}(H)$  observed at high magnetic fields is in agreement with the hysteresis observed in electric polarization experiments~\cite{Kimura_PRL_2009}. The electric field that is necessary for polarization of the sample increases with the  magnetic field. Therefore the change of polarization of the sample by the permanent electric field $E_{\_}=500$~kV/m occurs only at magnetic fields below $~5$~T.

\begin{figure}[h]
\includegraphics[width=0.95\columnwidth,angle=0,clip]{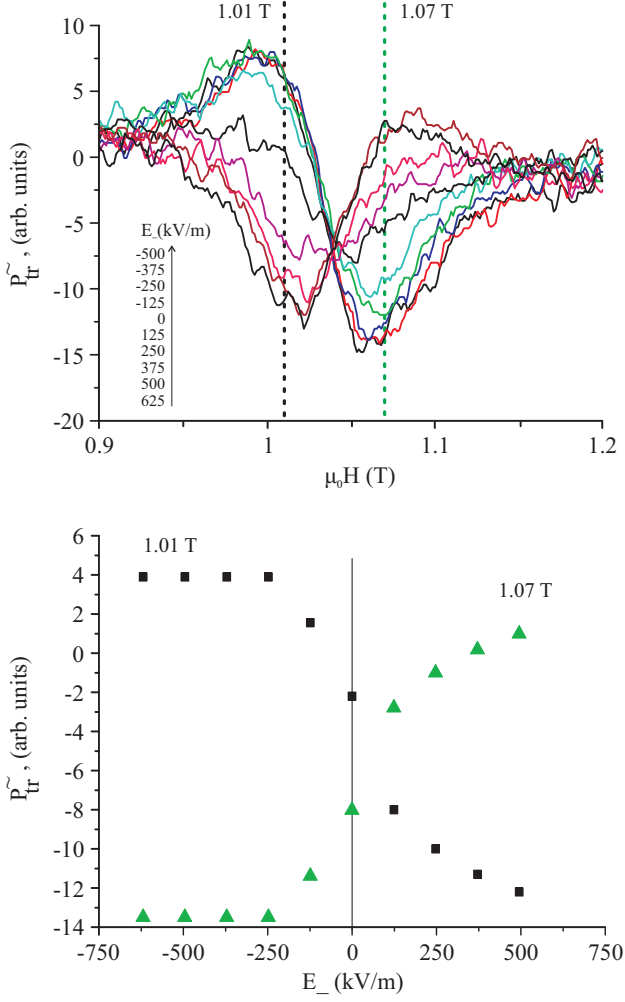}
\caption{(color online) Upper panel: $P^{\sim}_{tr}(H)$ measured at $\nu=42.2$~GHz, $T=4.2$~K and different values of the permanent electric field $E_{\_}$. $E^{\sim}=250$~kV/m. Bottom panel: $P^{\sim}_{tr}(E_{\_})$ measured at $H=1.01$, $1.07$~T.
}
\label{fig:line3}
\end{figure}

The upper panel of Fig.~\ref{fig:line3} shows $P^{\sim}_{tr}(H)$ measured in the field range of the absorption line from domain ``A" at different values of the permanent field $E_{\_}$. The field scan of $P_{tr}^{\sim}$ has the shape of a distorted field derivative of the absorption line. The bottom panel of the figure shows the dependence of the amplitude of $P^{\sim}_{tr}$ on the permanent electric field $E_{\_}$ at fields H near the extrema shown in the upper panel with arrows. The figure demonstrates that the shape of $P^{\sim}_{tr}(H)$ at fields higher than  $E_{\_}=500$ kV/m saturates, which means that this field is sufficient for electrically polarizing of the sample.

\begin{figure}[h]
\includegraphics[width=0.95\columnwidth,angle=0,clip]{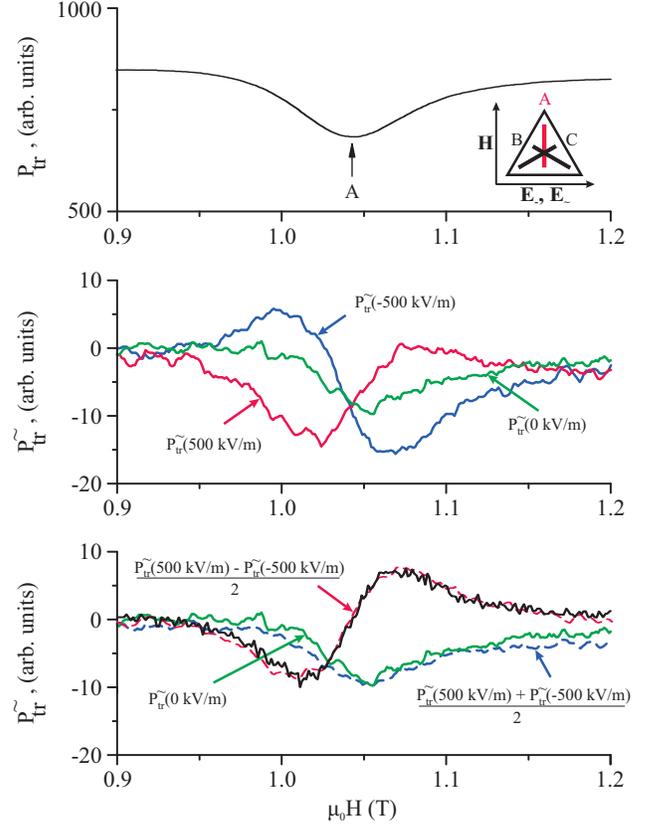}
\caption{(color online) Upper panel: Field scan of $P_{tr}(H)$. Middle panel: $P^{\sim}_{tr}(H)$ measured at the permanent electric fields $E_{\_}=0$ and $E_{\_}=\pm 500$~kV/m. $E^{\sim}=250$~kV/m. Bottom panel: field scans of the algebraic half-sum and half-difference of  $P^{\sim}_{tr}(H)$ measured at $E_{\_}=+500$~kV/m and $E_{\_}=-500$~kV/m (blue and red dotted lines). The algebraic sum is close to  $P^{\sim}_{tr}(H)$ measured at $E_{\_}=0$ (green solid line). The black solid line shows a scaled field derivative of the transmitted power. $\nu=42.2$~GHz and $T=4.2$~K.
}
\label{fig:line4}
\end{figure}

The upper and middle panels of Fig.~\ref{fig:line4} show the field scans of $P_{tr}(H)$ and  $P^{\sim}_{tr}(H)$ in the range of the low-field absorption line from domain ``A" measured at the permanent electric fields $E_{\_}=0$ and $E_{\_}=\pm 500$~kV/m. The bottom panel shows the algebraic half-sum and half-difference of  $P^{\sim}_{tr}(H)$ measured at $E_{\_}=\pm 500$~kV/m given by dotted lines. The algebraic half-sum is close to  $P^{\sim}_{tr}(H)$ measured at $E_{\_}=0$, whereas the difference is described well by a scaled field derivative of the transmitted power ($dP_{tr}/dH$). The half-difference line obtained with such a procedure was reproducible, whereas the half-sum was dependent on the preparation method of electrodes and the cooling history of the sample. The half-sum $(P^{\sim}_{tr}(E_{\_}=+500~{\rm kV/m})+P^{\sim}_{tr}(E_{\_}=-500~{\rm kV/m}))/2$ can be well fitted by a scaling of the absorption line $P_{tr}(H)$.

\begin{figure}[h]
\includegraphics[width=0.95\columnwidth,angle=0,clip]{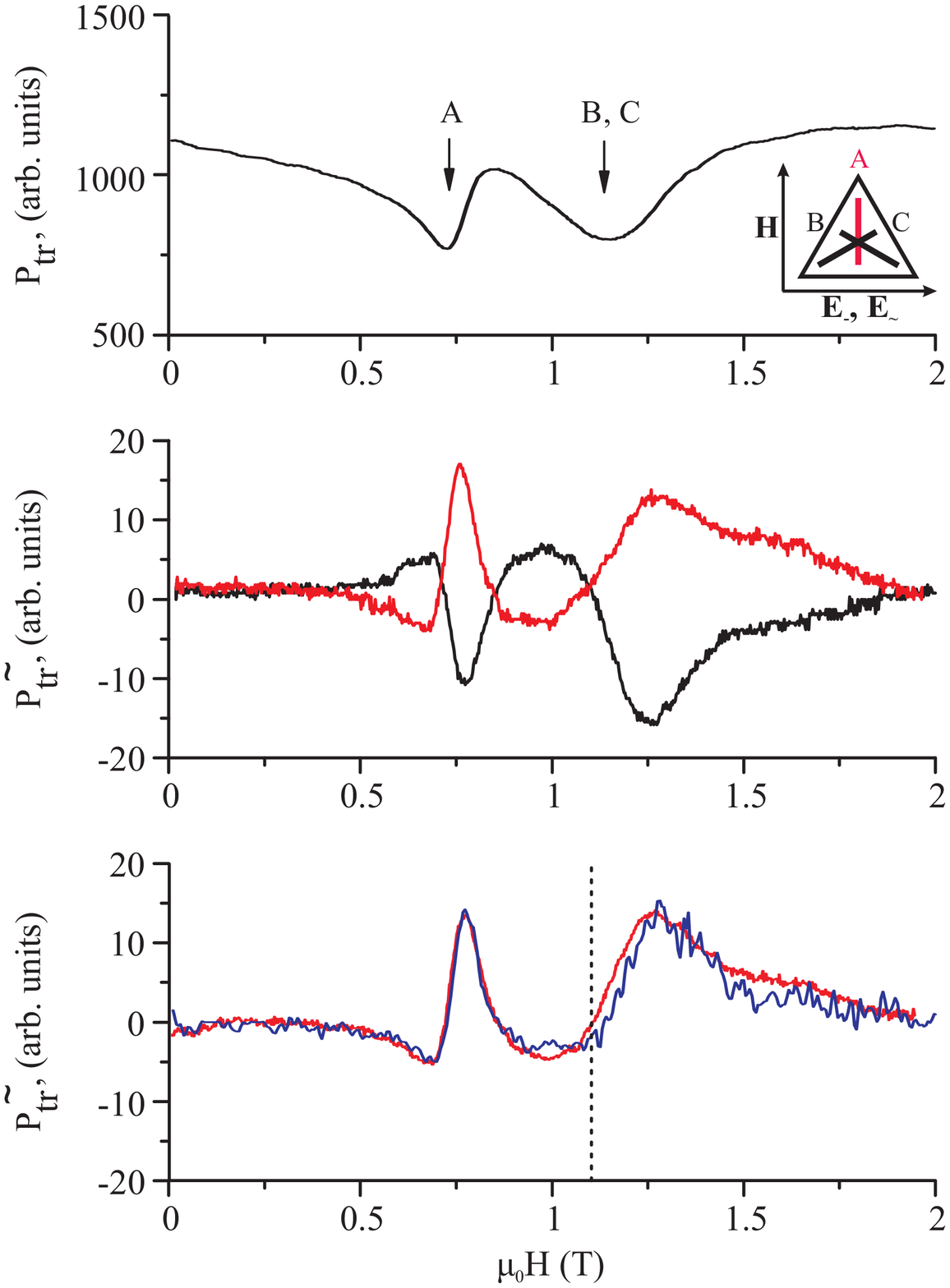}
\caption{Upper panel: Field scan of $P_{tr}$. Middle panel: $P^{\sim}_{tr}(H)$ measured at $E^{\sim}=125$~kV/m and $E_{\_}=+500$~kV/m, $E_{\_}=-500$~kV/m (red and black lines, respectively). Bottom panel: field scan of the half-difference of  $P^{\sim}_{tr}(H)$.
The blue solid line shows a scaled field derivative of the transmitted power. Two different coefficients were used for scaling: one for $H<1.1$~T and the other for higher fields. $\nu=38.6$~GHz and $T=4.2$~K.
}
\label{fig:line5}
\end{figure}

Figure ~\ref{fig:line5} demonstrates the field scans of $P_{tr}(H)$ and  $P^{\sim}_{tr}(H)$ measured at $E_=\pm 500$~kV/m,  $\nu=38.6$~GHz and $T=4.2$~K. In this measurement, the half-sum signal was accidentally small. In the bottom panel, the half-difference of $P^{\sim}_{tr}(H)$ measured at $E_{\_}=\pm 500$~kV/m and $\nu=38$~GHz is shown. The line can be fitted by a scaling of the field derivative of the transmitted power. Two scaling coefficients for the fit of this line were used. For fields of the low-field absorption line from domain ``A" ($\mu_0 H < 0.85$~T), the scaling coefficient was approximately two times smaller than the coefficient for the higher-field range, where the absorption line from domains ``B" and ``C" was observed.

\section{Discussion}

The oscillating response of the high-frequency power $P^{\sim}_{tr}(H)$ transmitted through the resonator on the applied oscillating electric field  in an electrically polarized sample can be divided to two parts. One part is proportional to the field derivative of the transmitted power and the second part reproduces the shape of absorption lines. The first part can be ascribed to the response caused by a linear shift of the absorption line by the applied electric field, whereas the second part can be ascribed to a change of the intensity of resonance absorption by $E^{\sim}$. The first part was well reproducible and can be described in the framework of a theoretical model, whereas the second contribution was dependent on the cooling history and on the technology of fabrication of the electrode. Presumably, the second uncontrolled part of the response is connected with a change in the magnetic domain distribution of the CuCrO$_2$ sample by the alternating electric field. In what follows, we discuss the expected response $P^{\sim}_{tr}(H)$ connected with the shift of the absorption line by the applied electric field. Because the shift of the resonance field $H_R$ in the electric field is small, the change in transmitted power can be written as

\begin{eqnarray}
P_{tr}(H,E\_+E^\sim)-P_{tr}(H,0)=\\ \nonumber \frac{\partial P_{tr}(H)}{\partial H}\cdot\frac{\partial H_{R}}{\partial E}\cdot (E\_+E^\sim),
\label{eqn:shift}
\end{eqnarray}
where the amplitude of the alternating electric field $E^\sim$ is defined by the applied alternating electric voltage divided by the distance between electrodes in the case of an electrically polarized sample.
The ESR frequency $\nu_3(H, E)$ in the framework of the model energy in Eq.~(\ref{eqn:energy}) for magnetic domain ``A" at the experimental orientations of fields is given by

\begin{eqnarray}
 H<H_c:\ \ \     \nu=\gamma\sqrt{\frac{\beta_2+\lambda_\perp E}{\chi_{\perp}}+H^2} \\ \nonumber
 H>H_c:\ \ \     \nu=\gamma\sqrt{-\frac{\beta_2+\lambda_{\perp} E}{\chi_{\perp}}+\frac{\chi_\parallel-\chi_\perp}{\chi_\perp}H^2}
\label{eqn:Espectra}
\end{eqnarray}
In this equation, the positive sign of $E$ corresponds to the case of the electric field codirected with polarization.
Combining Eqs.~(\ref{eqn:shift}) and (\ref{eqn:Espectra}) we obtain the expected value of $P^{\sim}_{tr}(H)$:

\begin{eqnarray}
P^\sim_{tr}=\frac{\partial P_{tr}(H)}{\partial H}\cdot\frac{ \lambda_\perp}{2  \chi_\perp H_{R}} E^\sim
\label{eqn:last}
\end{eqnarray}

It follows from this equation that the $P^\sim_{tr}$ in the framework of the model is expected to be:\\
i. independent on the sign of the magnetic field $\vect{H}$;\\
ii. proportional to the amplitude of $E^{\sim}$;\\
iii. dependent on the sign of electric polarization.\\
These specific features are demonstrated experimentally (see Figs. \ref{fig:line2}--\ref{fig:line5}).

The lines $P_{tr}(H)$ and $P^\sim_{tr}(H)$ in Figs.~\ref{fig:line1},~\ref{fig:line2},~\ref{fig:line4} and ~\ref{fig:line5} were measured in arbitrary, but the same units, which allows experimentally determing the absolute value of $\lambda_\perp$ that defines the spontaneous electric polarization of the sample.  The value of the electric polarization of CuCrO$_2$ obtained by processing the experimental results presented in Figs.~\ref{fig:line2}--\ref{fig:line5} for domain ``A" is $p= 110 \pm 15$~$\mu$C/m$^2$. This value is close to the values obtained by measuring the pyro-current, $p \approx 120$~$\mu$C/m$^2$.~\cite{Kimura_PRL_2009, Kimura_PRB_2008}.

The shift of the ESR absorption line from domains ``B" and ``C" is associated not only with the change of the energy gap of the ESR branch but also with the rotation of the spin plane in the electric field. The spectra $\nu(H,E)$ for these domains were computed numerically. The value of the polarization obtained from $P_{tr}^{\sim}(H)$ corresponding to the absorption line from domains ``B" and ``C" is $p=65\pm15$~$\mu$C/m$^2$. This underestimated value is most probably connected with the hysteretic spin plane oscillation in the alternating electric field $E^\sim$. The electric polarization evaluated in the case of the fully pinned spin plane (i.e., without taking the spin plane rotation in $E^\sim$ into account) is $p=160~ {\rm\mu C/m^2}$.  The value of the electric polarization  obtained from the pyro-current experiments is between the values obtained in the models of not pinned and fully pinned spin plane.

\section{Conclusions}
The effect of a permanent electric field on the ESR frequency was studied in the multiferroic CuCrO$_2$. It was shown that the observed shift of the ESR absorption line is independent of the sign of the static magnetic field and is proportional to the value of the applied electric field. The sign of the shift of the ESR absorption line depends on the electric polarization of the sample. The observations are in qualitative and quantitative agreement with the theoretical expectations.~\cite{Marchenko_2014}

This agreement with theory observed in the low-field spiral 3D-ordered magnetic phase in CuCrO$_2$ gives the hope that in the high-field range, where, according to NMR, the usual 3D magnetic order is destroyed, the observed electric polarization is indeed explained by 3D ordering of the chirality vector $\vect{n}$.\cite{Sakhratov_2016}

\acknowledgements

We thank V.~I.~Marchenko and A.~I.~Smirnov for the stimulating discussions. H.D.Z. acknowledges support from  NSF-DMR through the award DMR-1350002. This work was supported by RFBR (Grant 16-02-00688) and by RAS Presidium Program "Actual problems of low temperature physics".

\appendix
\section*{Appendix: The angular dependences of ESR without electric field}
To verify the model given by first three terms of Eq.~(2) (main text), the angular dependences of resonance field $H_R$ were studied.

The magnetic structure of CuCrO$_2$ without an electric field was studied previously with the AFMR technique for magnetic fields applied along the rational crystal axes: $\vect{H}\parallel$ [110], $[1\bar{1}0]$.~\cite{Yamaguchi_2010, Svistov_2013} To be sure, that the model of energy given by first three terms in Eq.~(2) (main materials) is adequate, the angular dependences were studied at $\nu = 42.25$~GHz within the $(110)$ and $(1\overline{1}0)$ planes and at three frequencies (36.1, 79.3 and 117.5~GHz) within the $(001)$ plane.

\begin{figure}[t]
\includegraphics[width=0.95\columnwidth,angle=0,clip]{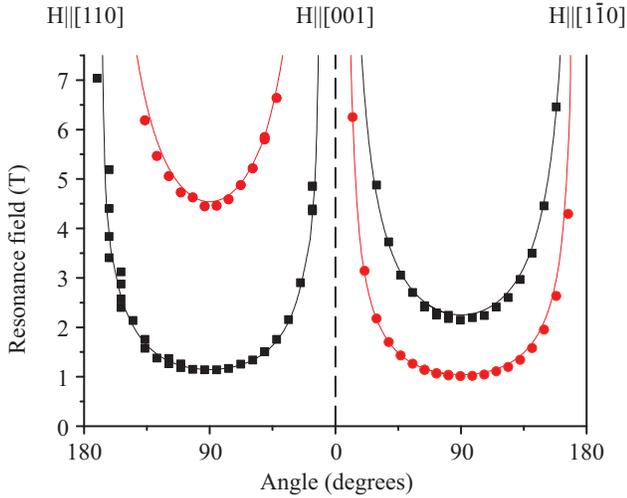}
\caption{(color online) The angular dependence of the resonance field $H_R(\alpha)$ for field $\vect{H}$ applied in the (110) and $(1\bar{1} 0)$ planes. Symbols: red circles correspond to the absorption lines from domain ``A", black squares - from domains ``B" and ``C". Lines: expected dependences: $H_R(\alpha)=H_R(\alpha=0)/cos(\alpha)$. $\nu=42.25$~GHz and $T=4.2$~K.
}
\label{fig:angle1}
\end{figure}

\begin{figure}[h]
\includegraphics[width=0.95\columnwidth,angle=0,clip]{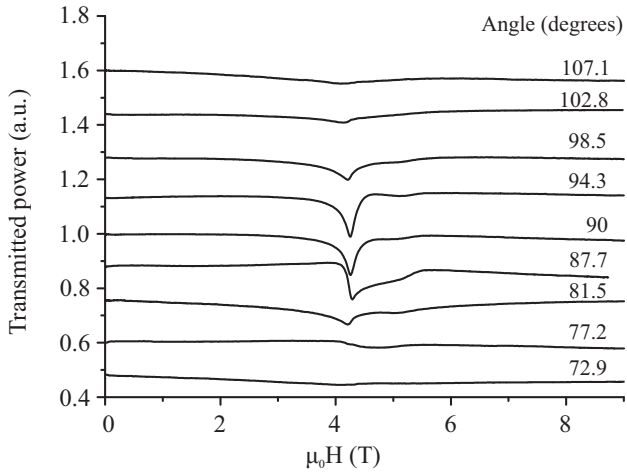}
\caption{(color online) Angular evolution of the field dependence records of the transmitted power at $\nu=$ 117.5~GHz and $T=1.3$~K. $\vect{H}$ rotates in the triangular plane.  $\alpha=0$ corresponds to the direction $\vect{H}\parallel$[110] for one of domains. The maximal intensity the transmitted power is normalized to one. Lines are shifted along the ordinate around line at $\alpha=90^\circ$ for clarity.
}
\label{fig:angle3}
\end{figure}

\begin{figure}[h]
\includegraphics[width=0.95\columnwidth,angle=0,clip]{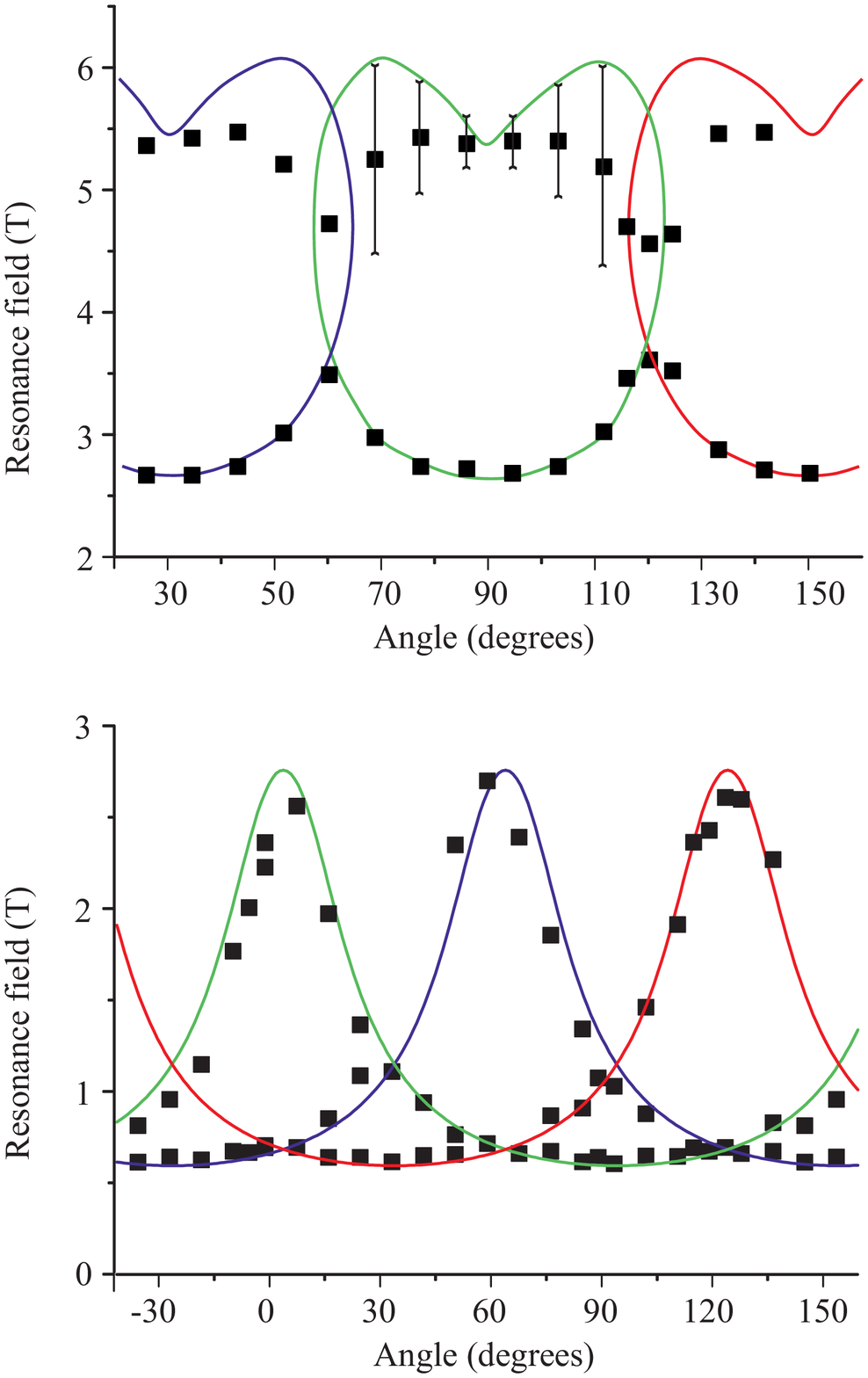}
\caption{(color online) Angular dependences of the resonance fields measured at $\nu=$ 79.3~GHz, T=1.3~K (symbols in the upper panel) and at $\nu=$ 36.1~ GHz and $T=4.2$~ K measured in Ref.~\onlinecite{Svistov_2013} (symbols in the bottom panel). $\vect{H}$ rotates in the $ab$ plane. Computed dependences of $H_R(\alpha)$ for three domains are given by solid lines of different colors.
}
\label{fig:angle4}
\end{figure}

Anisotropy perpendicular to the triangular plane is strong, so it is expected that the vector $\vect{n}$ practically does not deflect from the triangular plane. As a result, the resonance frequency $\nu_3(H)$ is expected to be defined by the projection of applied field $\vect{H}$ on the triangular $ab$ plane. To verify this, the angular dependences of resonance fields $\mu_0H_R$ were studied within the $(110)$ and $(1\overline{1}0)$ planes.  Fig.~\ref{fig:angle1} shows the dependences of the resonance fields $H_R$ (circles) on the angle $\alpha$ between $\vect{H}$ and the $ab$ plane. The observed angular dependences are well described by $H_R(\alpha=0)/cos(\alpha)$ (solid lines) for rotation of $\vect{H}$ in the $(110)$ and $(1\bar{1}0)$ planes.

The frequency--field diagram of antiferromagnetic resonances is expected to be strongly dependent on the direction of the applied field $\vect{H}$ within triangular plane (see \ref{fig:setup}). In accordance with the expectation, the absorption lines corresponding to one domain were observed in narrow angle range ($\approx 90\pm 8^\circ$)  at $\nu~=~117.5$~GHz and $T = 1.3$~K  (see Fig.~\ref{fig:angle3}). In Fig.~\ref{fig:angle4}, the experimental angular dependences of the resonance fields measured at $\nu=79.3$~GHz and $T=1.3$~K and at $\nu=36.1$~GHz and $T=4.2$~K are given with points and the theoretical dependences are given with solid lines. The theory basically agrees with the experiment except the range of the fields and angles, where the spin plane rotates on a large angle. In this angle--field range, the observed absorption lines are broad and nonsymmetric. This result, probably, indicates the distribution of the anisotropy parameter  $\beta_2$ in the sample. To avoid this complication, the influence of the electric field on ESR was studied mostly at $\vect{H}\parallel [1 \bar{1} 0]$  i. e. $\alpha =90^\circ$ and at fields $H\ll H_c$. In this case, the absorption line from domain ``A" is narrow and well separated from the absorption lines from two other domains, ``B" and ``C".

\bibliography{BibfileCuCrO2}

\end{document}